\documentclass[letter]{aa}
\usepackage{graphicx}
\graphicspath{{./}{Data/}}

\usepackage[T2A]{fontenc}
\usepackage[utf8]{inputenc}
\usepackage[english]{babel}
\usepackage[normalem]{ulem}  
\usepackage{xcolor}
\usepackage{multirow}
\usepackage{txfonts}
\begin{document}

	\title{Shock-wave heating mechanism of the distant solar wind: explanation of Voyager-2 data}
	\titlerunning{Shock-wave heating mechanism}
	
	
	\author{S.D. Korolkov
		\inst{1,2,3}\thanks{\email{korolkov.msu@mail.ru}}
		\and
		V.V. Izmodenov\inst{1,2,3}\thanks{\email{izmod@iki.rssi.ru}}
	}
	
	\institute{Space Research Institute (IKI) of Russian Academy of Sciences, Moscow, Russia\\
		\and
		Lomonosov Moscow State University, Moscow center for fundamental and applied mathematics, Moscow, Russia\\
		\and HSE University, 20 Myasnitskaya Ulitsa, Moscow 101000, Russia \\
	}
	
	\abstract 
	{} 
	{ One of the important discoveries made by Voyager-2 is the nonadiabatic radial profile of the solar wind proton temperature. This phenomenon has been studied for several decades. The dissipation of turbulence
energy has been proposed as the main physical process responsible for the temperature profile. The turbulence is both convected with the solar wind and originated in the solar wind by the compressions and shears in the flows and by pick-up ions. The compression source of the solar wind heating in the outer heliosphere appears due to shock waves, which originated either in the solar corona or in the solar wind itself.
        The goal of this work is to demonstrate that the shock-wave heating itself is enough to explain the temperature profile obtained by Voyager-2. 
	}
	{ The effect of shock-wave heating is demonstrated in the frame of a very simple spherically symmetric high-resolution (in both space and time) gas-dynamical data-driven solar wind model. This data-driven model employs the solar-wind parameters at 1 AU with minute resolution. The data are taken from the NASA OMNIWeb database. It is important to underline that (1) the model captures the shocks traveling and/or originating in the solar wind, and (2) other sources of heating are not taken into account in the model. We extended this simple model to the magnetohydrodynamic (MHD) and two-component models and found very similar results.}
	{The results of the numerical modeling with the one-minute OMNI data as the boundary condition show very good agreement with the solar-wind temperature profiles obtained by Voyager-2. It is also noteworthy that the numerical results with daily averaged OMNI data show a very similar temperature profile, while the numerical runs with 27-day-averaged OMNI data demonstrate the adiabatic behavior of the temperature. }
	{}

	
	\maketitle
	%
	
\section{Introduction}

Voyager-2 observations clearly demonstrated \citep{Gaziz1994, Lazarus1995} that the solar wind proton temperature does not decrease  with heliocentric distance adiabatically, as $T \sim 1/r^{2 \gamma -2}$ $(\gamma = 5/3$ for monoatomic gas as the solar wind). Instead, the temperature falls slowly out to $\sim$ 20-25 AU and  increases slowly at larger distances (see, e.g., Fig.~\ref{fig1}).

Most studies of the radial evolution of solar wind in the outer heliosphere explain the solar-wind heating by dissipation of the energy of turbulent fluctuations on small scales. As part of this approach, the turbulence transport equations are solved together with the solar-wind gas dynamic or magnetohydrodynamic (MHD) equations averaged over small timescales, which allows the effects of turbulence to be included in the right part of the energy equation. The turbulence transport model for solar-wind fluctuations was developed by \cite{Zhou1990} and \cite{Marsch1989}. The fluctuations are transported convectively with the solar wind and also generated by different mechanisms. It is necessary to note that in the absence of sources of turbulence, the fluctuations would rapidly decay and would not produce the observed heating. The generation of turbulent fluctuations is mainly associated with (1) a shear source related to transversal motion in the plasma \citep{Coleman1968}, (2) a compression source related to the shock wave \citep{Whang1991}, and (3) low-frequency MHD waves that arise during pickup origination and future isotropization of the pickup velocity distribution (see, e.g., \cite{Williams1995}). The compression source was introduced into the turbulence transport model of the solar wind for the first time by \cite{Zank1996}.

\cite{Smith2001} and \cite{Isenberg2003} also studied the turbulent transport of solar wind fluctuations and the role of their dissipation in the heating of thermal protons. The authors found that the radial temperature behavior correlates with observations for some values of the governing parameters. Further application of such an approach can be found in later works (\cite{Isenberg2010, Oughton2011, Gamayunov2014, Adhikari2014, Adhikari2021}).

The three-fluid 3D model, which includes turbulent transport, turbulent viscosity, turbulent resistivity, and turbulent heating, is presented in the study of \cite{Usmanov2014}. The results of modeling show good agreement with plasma observations by Wind, Ulysses, and Voyager-2 spacecraft. \cite{Usmanov2014} concluded that turbulent viscosity can influence the plasma temperature in the heliosphere (6\% of proton heating is shown to take place at a heliocentric distance of about 10 AU).

For a more detailed study of plasma heating by a compression source, we refer to the work of \cite{Whang1991}. This author, in analyses of Interplanetary Monitoring Platform (IMP), Voyager, and Pioneer solar-wind data, carried out numerical simulations resolving the Rankine-Hugoniot conditions on the shock waves propagating in the solar wind and demonstrated that interplanetary shock waves are dominantly responsible for the solar wind plasma heating between 1 and 15 AU. The shock waves that appear repeatedly in the nearby solar wind are due to both propagations of the shocks created in the solar corona and the formation of the shocks in the wind. 
The strongest shock waves are mostly connected with coronal mass ejections (CMEs) propagating from the solar corona and corotating interaction regions (CIRs) arising in the interplanetary medium. A series of works were published concerned with propagating particular CMEs and CIRs in the heliosphere (see, e.g., \cite{Gazis_2000, Wang2004}). Also, there are other types of structures in the solar wind (see, e.g. \cite{Ermolaev2021}). Here, we do not examine the propagation of the individual structures using a different approach.

In this study, we follow the approach used by \cite{Whang1991}. As an alternative to the turbulence transport model with the heating source terms in the energy equation, we solve time-dependent ideal gas-dynamic equations with as much time and space resolution as possible. It should be emphasized that we do not have any source terms in the equation to heat the solar wind (except for model 3). We used the boundary conditions based on OMNI data (from 1978 to 2005). In the terms of the turbulence transport model, our model includes (without any assumptions) the heating by shock waves. Other mechanisms are not taken into account.

It should also be emphasized that (1) our model is data driven and based on the high-resolution solar wind at 1 AU obtained from the OMNI database, and that (2) the model does not have any ad hoc assumptions.

The structure of this paper is as follows: In Sect.~\ref{glav1} we briefly discuss the models used in this work, and in Sect.~\ref{glav2} we show the results of calculations and their comparison with the Voyager-2 data. In Sect.~\ref{glav3} we discuss and summarize these results. Also, Appendix~\ref{ap1} contains a detailed description of the models, Appendix~\ref{ap2} contains a simple demonstration of shock-wave heating for a single shock layer, Appendix~\ref{ap3} contains additional calculation results (e.g., not averaged over time), and Appendix~\ref{ap4} demonstrates the ability of models to catch the shock wave at 1 AU for various resolutions of the boundary conditions.

\section{Models}
\label{glav1}

All the main results and conclusions presented in this paper can be obtained in the framework of a very simple model, referred to as Model 1 hereafter.
In the framework of this model, we solve nonstationary 1D Euler equations for ideal gas dynamics. Spherical symmetry is assumed.
The solar wind protons and electrons are considered as one fluid. The interplanetary magnetic field is not taken into account.
It is highly important that the numerical code is able to capture shock waves, which necessitates solving the Euler equations in a conservative form, which preserves the mass, momentum, and energy flows at the discontinuities. We used a Godunov-type scheme with linear interpolation within the cells. To avoid numerical smoothing of the results, we use the high-resolution grid with 320\,000 cells from 1 to 80 AU. We also verified that a larger number of cells does not influence the results. It is important to note that we additionally included magnetic fields in the model, but this does not lead to noticeable changes in the results (see Appendix~\ref{prilA}), and therefore only the gas-dynamic model is presented in the core of the paper.

A distinctive feature of our models is nonstationary boundary conditions. At 1 AU, OMNI data for the velocity, density, and temperature of solar wind protons are used as internal boundary conditions. The inner boundary condition imposes additional restrictions on the time step in the numerical scheme, which was chosen to have a minimum of (1) the time step dictated by the standard Courant’s criterion, and (2) 0.1 of the selected temporal resolution of the OMNI data, which is used as the boundary conditions. There is no need to set the boundary conditions at the outer boundary because the flow is supersonic and the directions of all the characteristics of the hyperbolic equations are toward the boundary. The outer boundary is placed closer to the Sun than the termination shock, and so we consider only the supersonic wind.
 
In the modeling calculations, we employed OMNI data with 1min, 1h, 1d, and 27d time resolutions. We find (see Appendix~\ref{AppC}) that 
daily averaged data are sufficient to reproduce the discussed effect.  The calculations performed with higher resolution produce similar results.

In addition to Model 1, we present results obtained in the frameworks of two other, more complicated models.
Model 2 is the two-fluid model in which solar wind plasma (protons and electrons) and pickup protons are considered as two co-moving fluids. The main purpose of the model is to obtain deceleration of the solar wind due to pickup mass-loading and thereby obtain better agreement with the solar-wind speed measured by Voyager-2. Similar to Model 1, Model 2 does not have free ad hoc parameters.
It is important that no exchange of thermal energy between the two components is assumed in Model 2. 
Model 3 is almost the same as Model 2 but some redistribution of the thermal energy from pickup protons to the solar wind is allowed. In model 3, during charge exchange, 99\% of the thermal energy goes to the pickup component, and 1\% is transferred to the thermal component. A similar assumption was made in the work of \cite{Wang2001}, but these authors assumed that 5\% of the energy is transferred to thermal protons.
A brief summary of the models is given in Table~\ref{tab:models}. A more detailed description of the models is given in Appendix~\ref{prilA}.

\begin{table}
        \caption{Brief description of the models used in this work}
        \label{tab:models}
        \begin{tabular}{ccc}
                \hline
                Models\  & Charge exchange effect & Energy \\
                &    &  redistribution$^*$\\
                \hline
                Model 1 & $-$ & $-$\\
                \hline
                Model 2 & $+$ & $-$\\
                \hline
                Model 3 & $+$ & $+$\\
                \hline
                \multicolumn{3}{l}{$^*$ with $\alpha = 0.01$, see in Appendix~\ref{mod3}}  \\
        \end{tabular}
\end{table}

\section{Results}
\label{glav2}

\begin{figure*}
	\includegraphics[width=\hsize]{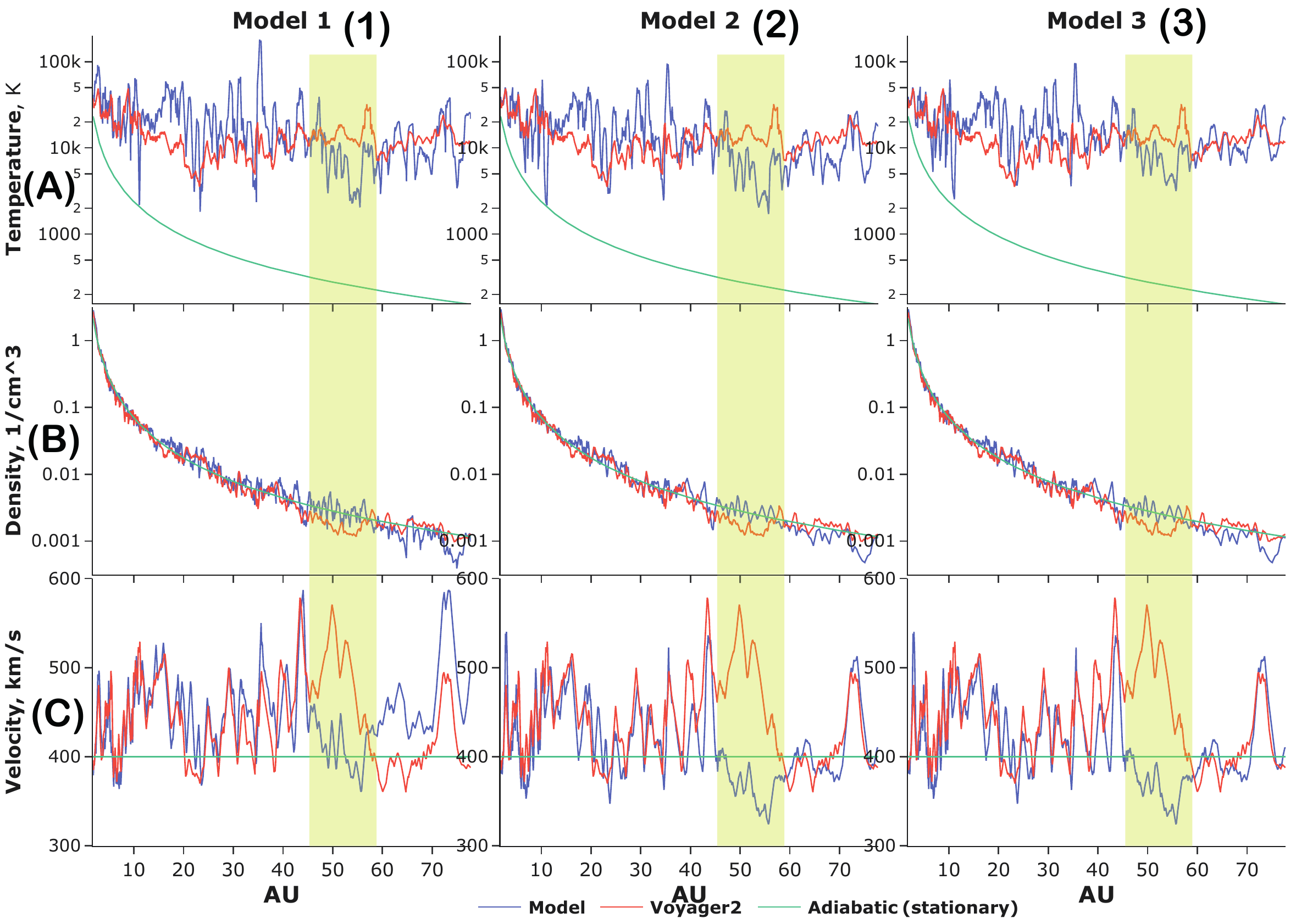}
	\caption{Fifty-day moving averages of temperature, density, and velocity of the solar wind obtained in the frameworks of  three models: Model 1 (first column), Model 2 (second column), and Model 3 (third column). Blue curves are the model results, red curves are Voyager-2 data, and green curves are the stationary solutions. $n_H = 0.05\  \text{1/cm}^\text{3}$. The yellow area corresponds to the high-speed solar wind from high heliolatitudes, and so this period cannot be compared with the models.}
	\label{fig1}
\end{figure*}

Figure~\ref{fig1} demonstrates (blue curves) the results of numerical calculations obtained in the frameworks of the three models discussed above. The distributions of the plasma parameters were calculated along the Voyager-2 trajectory. Models 1, 2, and 3 correspond to the left (1), central (2), and right (3) columns in the figure, respectively.  Voyager-2 data are shown in Fig.~\ref{fig1} as red curves.  The stationary adiabatic solution is shown by green curves.  In Fig.~\ref{fig1}, we plot 50d moving averages because nonaveraged results are very noisy. The timescale for averaging was chosen arbitrarily.  Nonaveraged results are presented in Appendix~\ref{pril1}. 

All models (including the simplest Model 1) can be clearly seen to reproduce the nonadiabatic behavior of the solar wind in the outer heliosphere. This is the main effect that we want to demonstrate in this paper.

The agreement with data provided by Voyager-2 is somewhat better for Models 2 and 3. Indeed, the bottom row (row C) of  Fig.~\ref{fig1} presents the distribution of the solar-wind velocity. The plot for Model 1 (panel 1C in Fig.~\ref{fig1}) clearly shows that the velocity measured by Voyager-2 is lower than that obtained in Model 1. This is the result of a well-known effect of solar wind deceleration due to the charge exchange of the solar wind protons with interstellar hydrogen atoms. Models 2 and 3 take this effect into account, and this is why we see good agreement between model and data in panels 2C and 3C of Fig.~\ref{fig1}. Here we should note that the effect of deceleration depends on the number density of the interstellar atoms. The value of the number density was chosen so as to obtain maximal agreement between model and data. We find that the best agreement is reached for 0.05 cm$^{-3}$ at distances of 70-80 AU. This value is very close to the value obtained by \cite{Richardson1995} and is lower than that obtained by \cite{Richardson2008}.

Therefore, we conclude that in Models 2 and 3, the radial profile of the solar wind speed agrees with Voyager-2 data exceptionally well, everywhere except the interval of distances between  $\sim$45 and $\sim$58 AU (shown as a yellow stripe in Fig. 1). This interval along the Voyager-2 trajectory corresponds to the years 1995-1999  of the solar minimum when Voyager-2 was located outside the sector zone \citep{Burlaga2003} in the high-speed streams of solar wind. Our spherically symmetric model is based on the OMNI data and comparison of the model results and Voyager-2 data is valid only when the spacecraft is inside the sector zone. This is why this period should be excluded from consideration.

Comparing Voyager-2 data with models for the solar wind proton number density, we find very good agreement for distances shorter than 45 AU. Beyond 60 AU, all models can be seen to have slightly smaller number density compared to Voyager-2 data. We do not know the exact reason for these small differences; they might be related to the deflection of the solar wind from the spherical symmetry, for example. 

Returning to the radial temperature profile, the results of Model 3 are shown in Fig.~\ref{fig1} (third column). In this model, we assume that only  1\% of the heat energy is redistributed from pickup protons to the solar wind. Comparing the results of Model 3 and Model 2, we see that 1 \% is enough to slightly increase the solar wind temperature, and therefore to obtain perfect agreement with Voyager-2 data at large (> 70 AU) distances.  Despite these exciting results, we underline that the main reason for the nonadiabatic behavior of the solar wind is not the heat-energy transfer from pickup protons. This effect is connected with the propagation of the shock waves and is discussed in the following section.

We also note two more interesting results obtained in the numerical calculations. The first is the higher solar wind temperature obtained in the models (all models) at distances of 15-40 AU compared to Voyager-2 data. At distances larger than 40 AU, the agreement between data and models becomes better.

For the second result, we refer to Fig.~\ref{fig2} (Appendix~\ref{ap3}), which shows nonaveraged daily model results and data. The levels of density and temperature fluctuations obtained in the framework of Model 1 are clearly seen to be larger than in the Voyager-2 data. However, once charge exchange is included in Models 2 and 3, the levels of these fluctuations reduce and become comparable with data (especially at large distances). To be more precise, the temperature minima found in the results of Model 1 become larger for Models 2 and 3.

\section{Discussion and conclusion}
\label{discas}
\label{glav3}

In this section, we discuss and interpret the results reported above. In light of our results, we are mainly concerned with the mechanism responsible for the nonadiabatic behavior of the solar wind obtained in all models including the simplest gas-dynamical Model 1. We believe this to be connected with the presence of numerous shocks in the solar wind and the behavior of the post-shock temperature with radial distance. 
First, let us consider a shock that propagates in the solar wind. The post-shock temperature, T, can be obtained from the Rankine-Hugoniot conditions as

\begin{equation} \label{eq_T}
	\frac{T}{T_*} = \dfrac{\left(1 + \dfrac{\gamma - 1}{2} M_*^2\right)\left(\dfrac{2\gamma}{\gamma - 1} M_*^2 - 1\right)}{M_*^2\left(\dfrac{2\gamma}{\gamma - 1} + \dfrac{\gamma - 1}{2}\right) },
\end{equation}

where $T_*$ and $\ M_*$ are the temperature and Mach number upstream of the shock.

If we assume a stationary adiabatic solution  $M_*^2 \sim r^{2\gamma - 2}$ and $T_* \sim r^{2-2\gamma }$ $(\gamma = 5/3)$ then, as follows from Eq. (\ref{eq_T}), the post-shock temperature $T$ approaches a constant (while $r \rightarrow \infty$).  In reality, at distances as short as 2-3 AU, the temperature is already almost constant (see the green curve in Fig.~\ref{T2}). Therefore, it is clearly seen that despite the temperature upstream of the shock falling down adiabatically as $T_* \sim r^{2-2\gamma }$, the temperature downstream of the shock remains constant with heliocentric distance. The contrast between upstream and downstream temperatures increases drastically with distance.
In order to obtain an average temperature (per day or per any other period) at any given distance, we performed averaging for both upstream and downstream temperatures. Further from the Sun, higher downstream temperatures affect the averaging.  In our view, this is the main reason for the observed nonadiabatic behavior of the solar wind.

Typically, single shock waves are not observed, but rather shock layers, limited to two shocks (forward and reverse).  Such structures are created, for example, in CIRs, which are produced twice per Sun rotation \citep{Whang1991}. The behavior of the temperature in such shock layers is explored in Appendix~\ref{ap2}. We show that the temperature is also nearly constant inside the single shock layer.  Moreover, the width of the shock layer increases with distance, which leads to an additional increase in the time-averaged temperature with distance. 

We refer to the physical effect of an increase in the average wind temperature due to the passage of the shocks or shock layers as the shock-wave heating mechanism of the solar wind. Our results clearly demonstrate that the effect of shock-wave heating almost completely explains the temperature distributions of solar wind protons measured by Voyager 2. To obtain better agreement with data, only 1\% of heat energy transfer was required from pickups in Model 3.

We also speculate that the higher temperature obtained in the models at 15-40 AU could be, for example, due to the slower formation of the shock waves and their interaction in the collisionless solar wind plasma, as compared with the ideal gas-dynamical solution.

Finally, we summarize our conclusions as follows:
\begin{itemize}
	\item  We show that the shock-wave heating mechanism described in this paper is the main mechanism that leads to the nonadiabatic behavior of the solar wind temperature observed by Voyager 2. The shock waves are generally seen in the solar wind and are generated in our numerical model constantly due to the strong variations in OMNI data at 1 AU that are used as the inner boundary conditions. It is sufficient to use one-day averaged boundary conditions to take into account the shock-wave heating effect in the nonstationary solar wind (see Appendix~\ref{ap3} and \ref{ap4} for details).
	\item The charge exchange effect with interstellar hydrogen atoms is important for slowing down the solar wind at large heliocentric distances. We introduced the two-fluid model (Model 2) of the solar wind in which the pickup component is co-moving with the core solar wind but is thermally decoupled. This model has no ad hoc parameters. In the framework of this model, we show that 0.05 cm$^{-3}$ is a sufficient interstellar H atom number density at 70-80 AU to obtain the deceleration observed by Voyager-2.
	\item Only 1\% of the thermal energy of the pickup component is required to be additionally transferred to thermal protons in order to achieve the observed heating in the outer wind.
\end{itemize}

The advantages of our model are that (1) it is data-driven, that is, all obtained results are based on the solar-wind parameters measured at 1 AU over a long period of time (from 1978 to 2005), and (2) it has no free parameters.

	\begin{acknowledgements}
		The work was performed in the framework of the Russian Science Foundation grant 19-12-00383.
	\end{acknowledgements}

	\bibliographystyle{aa}
	\bibliography{yourfile}

\begin{appendix}
\section{Models description}
\label{prilA}
\label{ap1}
This Appendix provides a detailed description of the models employed in the paper. 

\subsection{Model 1}
We start with the most simple, Model 1. As reported in Sect.~\ref{glav1}, this model describes the spherically symmetric supersonic flow of ideal gas. The solar wind is assumed quasi-neutral. A single fluid approach is employed. No interaction with the interstellar atoms is assumed.

The OMNIWeb database provides three components of the solar-wind velocity and three components of the heliospheric magnetic field. For the spherically symmetric problem,  we assume that $\dfrac{\partial }{\partial \varphi} = \dfrac{\partial }{\partial \theta} = 0$ for all velocity and magnetic field components with one exception. The derivative $\dfrac{\partial B_\theta }{ \partial \theta}$ cannot be considered as zero due to the condition $\mathrm{div} \mathbf{B} = 0$ for solenoidal magnetic field. Nevertheless, this derivative can be derived from the solenoidal condition.
The closed system of ideal MHD equations for a spherically symmetric problem can be written  in the spherical coordinate system as follows:

\begin{eqnarray} \label{eq_MHD_spherical}
	\begin{cases}
		\dfrac{\partial \rho}{\partial t} + \dfrac{1}{r^2} \dfrac{\partial}{\partial r} \left(r^2\rho V_r \right) = 0, \\[3mm]
		\dfrac{\partial \rho V_r}{\partial t} + \dfrac{1}{r^2} \dfrac{\partial}{\partial r} \left(r^2 (\rho V_r^2 + p^* - B_r^2) \right) - \dfrac{B_r}{r} \dfrac{\partial B_\theta }{\partial \theta} = \\[3mm]
		= \dfrac{\rho (V_\theta^2 + V_\varphi^2) + 2p^* - (B^2_\varphi + B^2_\theta)}{r}, \\[3mm]
		\dfrac{\partial \rho V_\theta}{\partial t} + \dfrac{1}{r^2} \dfrac{\partial}{\partial r} \left(r^2 (\rho V_r V_\theta - B_rB_\theta) \right) - \dfrac{B_\theta}{r} \dfrac{\partial B_\theta }{\partial \theta} = \\[3mm] 
		=\dfrac{B_rB_\theta-\rho V_r V_\theta}{r}, \\[3mm]
		\dfrac{\partial \rho V_\varphi}{\partial t} + \dfrac{1}{r^2} \dfrac{\partial}{\partial r} \left(r^2 (\rho V_r V_\varphi - B_rB_\varphi) \right) - \dfrac{B_\varphi}{r} \dfrac{\partial B_\theta }{\partial \theta} = \\[3mm] 
		= \dfrac{B_r B_\varphi - \rho V_r V_\varphi}{r}, \\[3mm]
		\dfrac{\partial B_r}{\partial t} - \dfrac{V_r}{r} \dfrac{\partial B_\theta }{\partial \theta} = 0, \\[3mm]
		\dfrac{\partial B_\theta}{\partial t} + \dfrac{1}{r^2} \dfrac{\partial}{ \partial r} \left( r^2(V_rB_\theta - V_\theta B_r)\right) = \dfrac{V_rB_\theta - V_\theta B_r}{r}, \\[3mm]
		\dfrac{\partial B_\varphi}{\partial t} + \dfrac{1}{r^2} \dfrac{\partial}{ \partial r} \left( r^2(V_rB_\varphi - V_\varphi B_r)\right) - \dfrac{V_\varphi}{r} \dfrac{\partial B_\theta }{\partial \theta}  = \\[3mm]
		= \dfrac{V_rB_\varphi - V_\varphi B_r}{r}, \\[3mm]
		\dfrac{\partial E}{\partial t} + \dfrac{1}{r^2} \dfrac{\partial}{\partial r} \left(r^2 \left[(E + p^*)V_r-B_r(\vec{B}\cdot\vec{V}) \right] \right) - \\[3mm]
		- \dfrac{(\vec{B}\cdot\vec{V})}{r} \dfrac{\partial B_\theta }{\partial \theta} = 0, \\[3mm]
		\dfrac{\partial B_\theta }{\partial \theta} = -r\dfrac{\partial B_r }{\partial r} - 2B_r,
	\end{cases}
\end{eqnarray}
where
\[
E = \frac{p}{\gamma-1}+ \frac{\rho {\vec{V}}^2}{2} + \frac{{\vec{B}}^2}{2}
\]
 is the total energy (we also note that we have chosen a system of units in which the magnetic permeability $\mu = 1$), and
\[
p^* = p + \frac{{\vec{B}}^2}{2}\ - \text{is the total pressure}.
\]

The last equation of the system is the condition of the solenoidal magnetic field and serves to determine $\dfrac{\partial B_\theta }{\partial \theta}$ at every moment of time and at every point in space.

\begin{figure*}
	\includegraphics[width=\hsize]{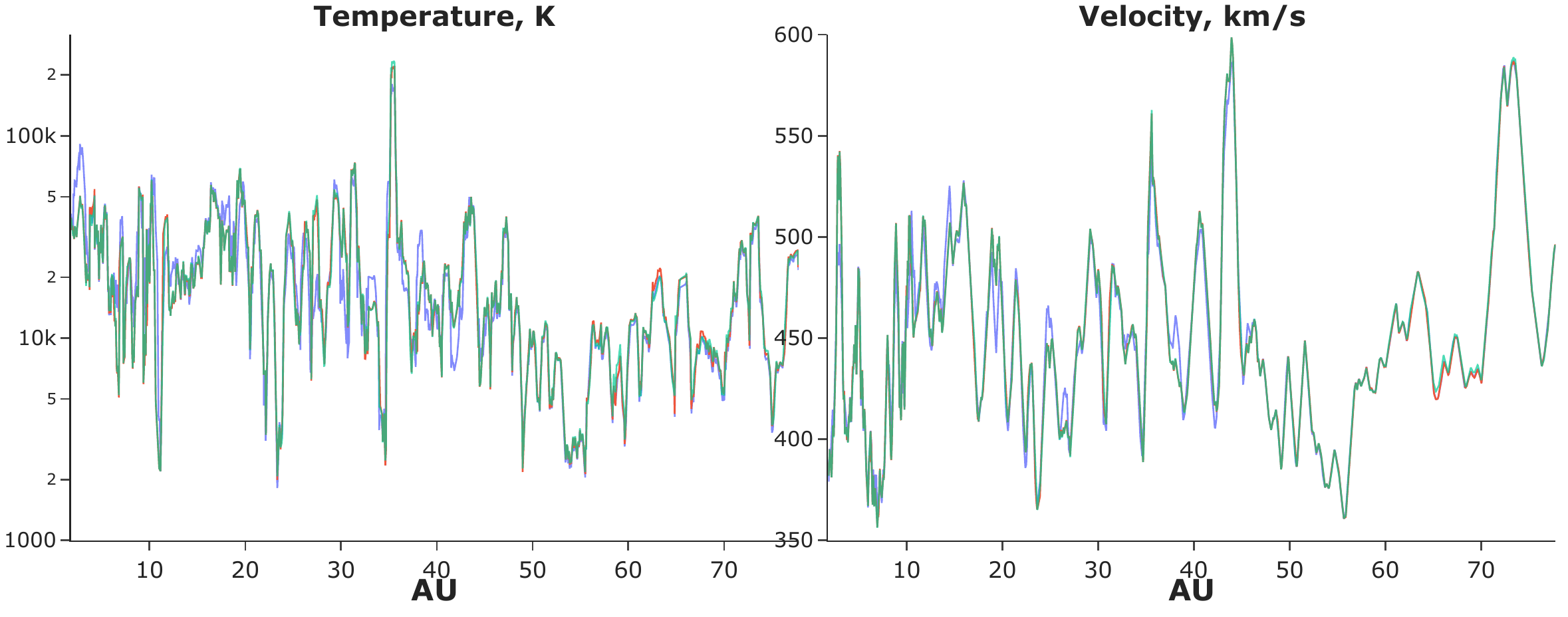}
	\caption{Fifty-day moving average temperature and velocity  distribution along the Voyager-2 trajectory. The results are presented for three model runs.
                Green curves correspond to the run in the framework of the model described by Eq. \ref{eq_MHD_spherical} with three nonzero components of the velocity and magnetic field.  Red curves correspond to the test run where the magnetic field is ignored. Blue curves correspond to the run where both magnetic field and nonradial components of the velocities are ignored. The figure shows that the results of all runs are identical and therefore a  simple 1D gas-dynamic model is sufficient for the purposes of this work.}
	\label{fig33}
\end{figure*}

In addition to the main calculations performed with all components of the velocity and the magnetic field, we performed two test runs. In the first test run, we assumed all components of the magnetic field to be zero. Therefore, in this run, we ignored the magnetic field completely and obtained a solution to the gas-dynamic equations. In the second test run we assumed that only the $V_r$ component of the solar wind velocity is not zero.  The results of all three runs (the main one and two test-runs) are shown in Fig.~\ref{fig33}.
It is seen that the difference between the three models is very minor.  Therefore, we conclude
that the gas-dynamic spherically symmetric model with only one nonzero velocity component is quite appropriate for our purposes. This model is employed in the main text of the paper as the base for Models 2 and 3.

\subsection{Models 2 and 3}
\label{mod3}
The main objective of introducing Models 2 and 3 is connected with the requirement to take into account the deceleration of the solar wind at large heliocentric distances due to charge exchange with the interstellar atoms. The deceleration is very well described in the frame of a one-fluid model (see, e.g., Fig.~4.3 in \cite{Izmodenov2006}). In the such a one-fluid model, the solar and pickup protons are indistinguishable. Their common temperature is close to the temperature of pickup protons because the temperature of the pickup proton is an order of magnitude higher than the temperature of the solar wind. Thus, the temperature of thermal protons cannot be correctly identified in a one-fluid model. Model 2 is the two-component model that provides the necessary deceleration and at the same time does not transfer the thermal energy from the pickup protons to the solar-wind protons.   Model 3 is an extension of Model 2 and allows some part of the thermal energy to be transferred to the solar wind protons.

Models 2 and 3 are very similar. They can be described by the system of equations (see Eqs. \ref{sys1}) with free parameter $\alpha$ in the right part of the energy equations. Here,
$\alpha$ determines the percentage of redistributed energy, and is zero for Model 2 and nonzero for Model 3. Below we describe the models in detail.

The governing equations of Model 2 are quite similar to those of previously developed two-component models (e.g., \cite{Zank2018}) but with some novelty. This is a two-fluid model in which thermal solar wind protons and pickup protons are considered separately. Pickup protons are formed as a result of the charge exchange of solar-wind protons with interstellar hydrogen atoms. To take into account the effect of charge exchange, two systems of Euler equations are solved together. One system is written for the mixture of all components (solar protons, pickup protons, and electrons), and the second system is for the solar wind protons and electrons. The speed of the solar wind protons is considered to be equal to the speed of the mixture as it follows from \cite{Isenberg1986}, who has shown that the bulk velocity of the newly created pickup protons relaxes to the bulk velocity of plasma much quicker as compared to the relaxation of the thermal velocities.



Model 2 is different from the previously known multi-component models (see, e.g., \cite{Wang2001} and \cite{Zank2018}). These models have five governing equations: (1-3) a conservative system of mass, momentum, and energy equations for the mixture of all co-moving components (solar wind protons, electrons, and pickups), (4) continuity equation for the pickup component, and (5) a heat flux equation (equation for internal energy) for the pickup component. Assuming that all components are co-moving, the five equations form the closed system of equations. However, this system of equations is not conservative, and so one cannot obtain Rankine-Hugoniot (R-H) conditions at the shocks. The R-H conditions are not needed in the papers by \cite{Zank2018} and \cite{Zank2014} because stationary (or quasi-stationary) supersonic solar wind flow is considered and the question about the R-H conditions at the shocks does not arise. Alternatively, our model is time-dependent and a correct modeling of the traveling shocks is critical for our purposes. To capture the shocks in our numerical code, we must solve a conservative system of equations.  Our system of equations consists of: (1-3) a conservative system of mass, momentum, and energy balance equations for all co-moving components (this is the same as in other models);  and (4-6) a conservative system of mass, momentum, and energy balance equations for only the solar wind component. Thus, our system resolves the relations at the shocks separately for the mixture of all components and for the thermal solar wind. In this case, the velocity jump at a shock for the mixture is different from the velocity jump for the thermal solar wind component. This is due to the different densities and pressures of the components at the shock front. Then, we want to satisfy a physically reasonable (and observable) condition that all components are co-moving. To do that, we have to add a source term in the momentum equation (term $Q_{2,\mathrm{p}}$ in the \ref{sys1} system) in order to equilibrate the velocities of the mixture and solar-wind components. From a physical point of view, this source is the force of interaction between the components, which accelerates the pickup component and slows down the thermal component of the solar wind, thereby ensuring the balance (equality) of velocities. The value of the source term can be easily obtained from the velocity equilibrium condition; we calculate it at each moment of time and then use it in the energy balance equation for the solar-wind component.

\begin{figure*}
	\includegraphics[width=\hsize]{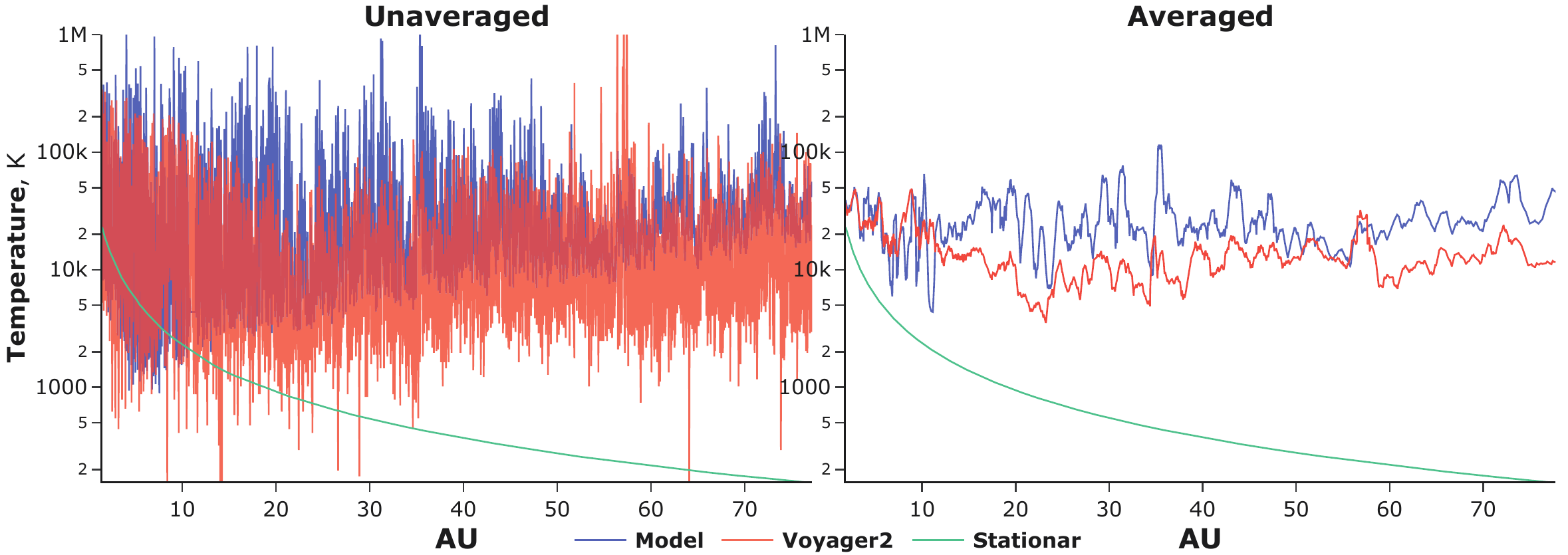}
	\caption{Radial profiles of the solar proton temperature obtained in the frame of Model 3 with $\alpha = 0.05$. The left panel corresponds to daily data as obtained in the model. The right panel corresponds to 50d moving averages.}
	\label{fig44}
\end{figure*}

As far as Model 3 is concerned, the possibility (as compared with Model 2) to redistribute part of the thermal energy between pickup protons and thermal protons is added in this model. Here we follow \cite{Wang2001},  who showed in the framework of their stationary model that 5\% of the thermal energy generated as a result of charge exchange should be transferred to thermal protons to reach the observed temperatures. In Model 3, 
an additional heat source was added to the solar wind energy balance equation in the system of Eqs.~\ref{sys1}. This approach makes it possible to add sources of thermal energy in addition to the shock-wave heating. In Sect.~\ref{glav2} of the paper, we demonstrate that 1\% of the redistribution is sufficient to make a better fit of Voyager 2 data.  Figure~\ref{fig44} demonstrates the temperature profile obtained in the framework of Model 3 with $\alpha = 0.05$, which corresponds to 5\% of the redistribution of the thermal energy.  It is seen that the model temperature, in this case, is sufficiently larger than observed by Voyager-2.

The system of the governing equations for Models 2 and 3 is the following:
\begin{eqnarray}
	\begin{cases}
		\dfrac{\partial \rho}{\partial t} + \text{div}( \rho\vec{V}) = 0,\\[3mm]
		\dfrac{\partial (\rho \vec{V})}{\partial t} + \text{div}( \rho\vec{V} \vec{V} + p\hat{I}) = \vec{Q}_2,\\[3mm]
		\dfrac{\partial E}{\partial t} + \text{div}( (E+p)\vec{V}) = Q_3,\\[3mm]
		\dfrac{\partial \rho_\mathrm{sw}}{\partial t} + \text{div}( \rho_\mathrm{sw}\vec{V}) = Q_{1,\mathrm{sw}},\\[3mm]
		\vec{Q}_{2,\mathrm{sw}} = \dfrac{\partial (\rho_\mathrm{sw} \vec{V})}{\partial t} + \text{div}( \rho_\mathrm{sw}\vec{V} \vec{V} + p_\mathrm{sw}\hat{I}),\\[3mm]
		\dfrac{\partial E_\mathrm{sw}}{\partial t} + \text{div}( (E_\mathrm{sw}+p_\mathrm{sw})\vec{V}) = \vec{Q}_{2,\mathrm{sw}} \cdot \vec{V} - \\[3mm]
		- Q_{1,\mathrm{sw}} \cdot \left( \dfrac{V^2}{2} - \dfrac{(\gamma + 1)p_\mathrm{sw}}{(\gamma - 1)\rho_\mathrm{sw}} \right) + \alpha \cdot Q_{3,\mathrm{sw}},\\[3mm]
	\end{cases}
	\label{sys1}
\end{eqnarray}
where
\[
E_\mathrm{sw} = \frac{p_\mathrm{sw}}{\gamma-1}+ \frac{\rho_\mathrm{sw} V^2}{2}, \\
E = \frac{p}{\gamma-1}+ \frac{\rho V^2}{2},
\]
\[
\rho = \rho_\mathrm{pui} + \rho_\mathrm{sw}, \\ p = p_{\mathrm{pui}} + p_\mathrm{sw}.
\]

Here, the quantities without an index denote the parameters of the mixture, those with the index $\mathrm{``sw}"$ denote the parameters of the solar wind, and $``\mathrm{pui}"$ denote the parameters of pickup protons. This system in the 1D (spherically symmetric) case consists of six equations and has six unknown parameters: $\rho,\ p,\ V,\ \rho_\mathrm{sw},\ p_\mathrm{sw},\ Q_{2,\mathrm{sw}} $. We note that the momentum equation of the thermal solar wind is used to determine the source term $Q_{2,sw}$. That is why we reverse the right and left parts of the equation as compared with the usual form.
Also, in the last equation of the system \ref{sys1}, the first term on the right side determines the contribution of the momentum source to the kinetic energy. The second term is related to the loss of total energy due to a decrease in mass of the thermal component due to charge exchange.

The expressions for source terms are written in the form of \cite{McNutt}:
\begin{eqnarray}
	\vec{Q}_2 & = & \nu_{\mathrm{H}}\cdot \left( \vec{V}_{\mathrm{H}} - \vec{V} \right) , \\ 
	Q_3 & = &  \nu_{\mathrm{H}}\cdot \left[  \dfrac{V^2_{\mathrm{H}} - V^2}{2}  + \dfrac{U^*_{\mathrm{H}}}{U^*_{\mathrm{M},\mathrm{H}}} (c^2_{\mathrm{H}} - c_\mathrm{sw}^2) \right],\\
	Q_{1,\mathrm{sw}} & = & - \frac{1}{m_\mathrm{p}} \rho_\mathrm{sw} \rho_{\mathrm{H}}  U^*_{\mathrm{H}} \sigma_{\mathrm{ex}}(U^*_{\mathrm{H}}), 
\end{eqnarray}
\begin{eqnarray}
	Q_{3,\mathrm{sw}}  =   \nu_{\mathrm{H}}\cdot \left[  \dfrac{|\vec{V}_{\mathrm{H}} - \vec{V}|^2}{2}  + \dfrac{U^*_{\mathrm{H}}}{U^*_{\mathrm{M},\mathrm{H}}} (c^2_{\mathrm{H}} - c_\mathrm{sw}^2) \right],
	\label{sys2}
\end{eqnarray}
\begin{eqnarray}
	\nu_{\mathrm{H}} & = & \frac{1}{m_\mathrm{p}} \rho_\mathrm{sw} \rho_{\mathrm{H}}  U^*_{\mathrm{M},\mathrm{H}} \sigma_{\mathrm{ex}}(U^*_{\mathrm{M},\mathrm{H}}),\\
	c_\mathrm{sw} & = & \sqrt{\dfrac{2k_\mathrm{B} T_\mathrm{sw}}{m_\mathrm{p}}},\ \ \ c_\mathrm{H} = \sqrt{\dfrac{2k_\mathrm{B} T_\mathrm{H}}{m_\mathrm{p}}}\\
	p_\mathrm{H} & = & n_\mathrm{H} k_\mathrm{B} T_\mathrm{H},\ \ \  p_\mathrm{sw}  =  2 n_\mathrm{p} k_\mathrm{B} T_\mathrm{sw},
\end{eqnarray}
where the index $``\mathrm{p}"$ denotes the parameters of thermal protons.

$\sigma_{\mathrm{ex}} (U)$ - charge exchange cross-section, depending on the speed.\\[1mm]
$\sigma_{\mathrm{ex}} (U) = (a_1 - a_2\cdot \mathrm{ln}(U))^2  $ in cm$^2$, where U in cm/s,  $a_1 = 1.64 \cdot 10^{-7}, a_2 = 6.95 \cdot 10^{-9} $\\

We note that $ Q_{3,\mathrm{sw}} \neq Q_3$, since it only includes thermal energy and does not include the work of forces.

The expressions for average speeds in the source term are as follows:
\begin{eqnarray}
	U^*_{\mathrm{H}} & = & \sqrt{|\vec{V}_{\mathrm{H}} - \vec{V}|^2 + \dfrac{4}{\pi}(c^2_{\mathrm{H}} + c_\mathrm{sw}^2)}\\[3mm]
	U^*_{\mathrm{M},\mathrm{H}} & = & \sqrt{|\vec{V}_{\mathrm{H}} - \vec{V}|^2 + \dfrac{64}{9\pi}(c^2_{\mathrm{H}} + c_\mathrm{sw}^2)}
\end{eqnarray}

The density of hydrogen is found from the solution of the continuity equation for hydrogen under the assumption of constant velocity of the hydrogen atoms within the heliosphere ($V_\mathrm{H} = 26.4$ km/s).
\begin{eqnarray}
	n_\mathrm{H} & = & n_{\mathrm{H},\infty} \cdot \exp(-r_0/r),\\
	r_0 & = & \dfrac{r_\mathrm{e}^2 v_\mathrm{e} \rho_\mathrm{e} \sigma_{\mathrm{ex}}}{m_\mathrm{p} V_{\mathrm{H}}}. \nonumber
\end{eqnarray}
The hydrogen temperature is also considered to be constant and equal to the temperature of the interstellar medium ($T_\mathrm{H} = 6527$ K). Therefore, $c_\mathrm{H}$ is constant. Also, $n_{\mathrm{H},\infty}$ is the parameter that will be chosen for the best match with the data.

\section{Shock-heating effect}
\begin{figure*}[htbp!]
	\centering
	\includegraphics[width=\hsize]{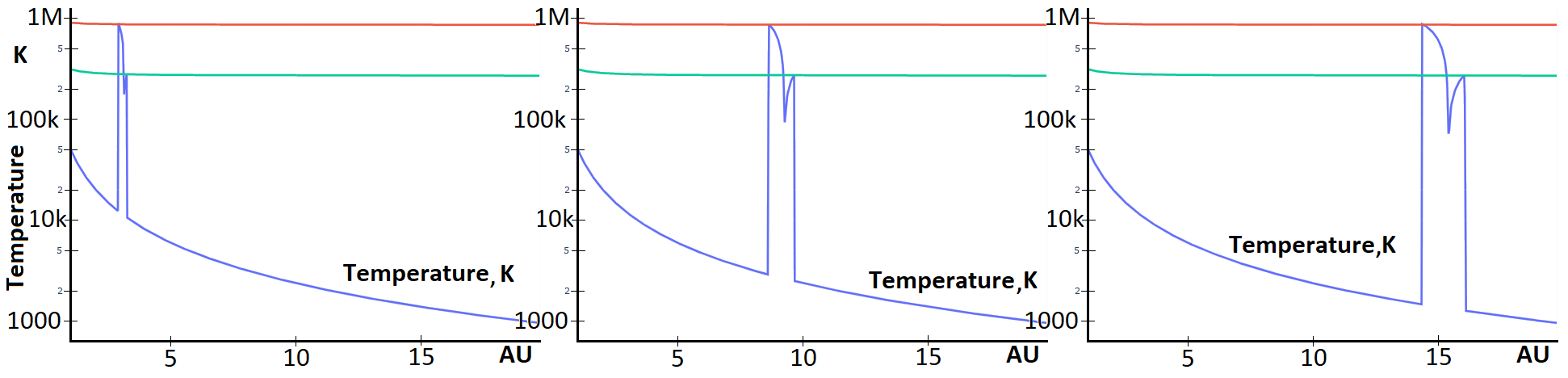}
	\caption{Temperature (blue curve) in a single shock layer as a function of the heliocentric distance presented at different time moments. The red and green curves represent an analytical estimate of the temperature jump on shock waves. The calculations were performed with the following parameters: $n_0 = 7\ \mathrm{cm}^{-3},\ V_0 = 375\ \mathrm{km/s},\ T_0 = 51109\ \mathrm{K} $.
	}
	\label{T2}
\end{figure*}

\label{effect}
\label{ap2}


In this Appendix, we explore the radial evolution of temperature in the shock layer consisting of forward and reverse shock.
The shock layer is formed by induced perturbation at the inner boundary ($r_0 = 1$ AU). We consider the following test problem.
At the initial moment of time ($t = t_0$), we assume the following radial distribution of the solar wind parameters, which corresponds to the stationary adiabatic solution of a hypersonic source:

\begin{eqnarray}
	n_1(r) = n_0 \left(\dfrac{r_0}{r}\right)^2, \ \ \ V_1(r) = V_0,\ \ \ T_1(r) = T_0 \left(\dfrac{r_0}{r}\right)^{2\gamma - 2}.
\end{eqnarray}
Then, for $t \ge t_0$ the inner boundary conditions are changed:
\begin{eqnarray}
	n_2 = \dfrac{n_0}{\chi^2}, \ \ \ V_2 = V_0 \cdot \chi,\ \ \ T_2 = T_0.
\end{eqnarray}
The dynamic pressure  $\rho V^2$ and temperature $T$ remain constant. The value of  $\chi$ was chosen to be equal to two in the presented calculations.

The solution of the Riemann problem at the inner boundary at the moment $t=t_0$ consists of two shocks and a tangential discontinuity between them (see \cite{Cherny}). This is how the shock layer is formed. Figure~\ref{T2} explores the evolution of the shock layer in the solar wind with time obtained in the numerical solution (blue curve). As the reverse shock moves more slowly than the forward shock, the width of the shock layer increases with time and distance. It is clearly seen from the numerical solution that the temperature inside the shock layer is nearly constant. This is explained by equation (\ref{eq_T}) because  the shock wave velocities quickly reach a constant value and are approximately $475\ \mathrm{km/s}$ and $529\ \mathrm{km/s}$ for the reverse and forward shocks (for $n_0 = 7\ \mathrm{cm}^{-3},\ V_0 = 375\ \mathrm{km/s},\ T_0 = 51109\ \mathrm{K},\ \chi = 2$), respectively. The analytically obtained temperatures behind the forward and reverse shocks are shown in Fig.~\ref{T2} by red and green curves, respectively. These temperatures are in good agreement with the numerical solution.

Therefore, the presented numerical solution demonstrates that when the shock layer propagates into the heliosphere (1) its width becomes larger, and (2) the temperature in the shock layer remains nearly constant. Both effects significantly increase the time-averaged temperature at a given point in space. 

This demonstration helps (in our opinion) to understand the results of our main simulations shown at the core of the paper.  To compare with the time-span-averaged Voyager-2 data, we have to average our numerical results with the same time spans. During a single time-span, different types of wind pass a given point. The types include shock layers (many of them since the boundary conditions are very inhomogeneous) and nonshocked solar wind that expands adiabatically. Our demonstration shows that the larger the heliocentric distance, the larger the width of the shock layers and the larger the difference in temperature between the adiabatic solar wind and shock-heated solar wind. In summary, over long distances, the high temperatures in the shock layers have a greater impact on the time-span-averaged temperatures that we have in the data.

\section{Additional calculation results}
\label{AppC}

\begin{figure*}
	\includegraphics[width=0.99\hsize]{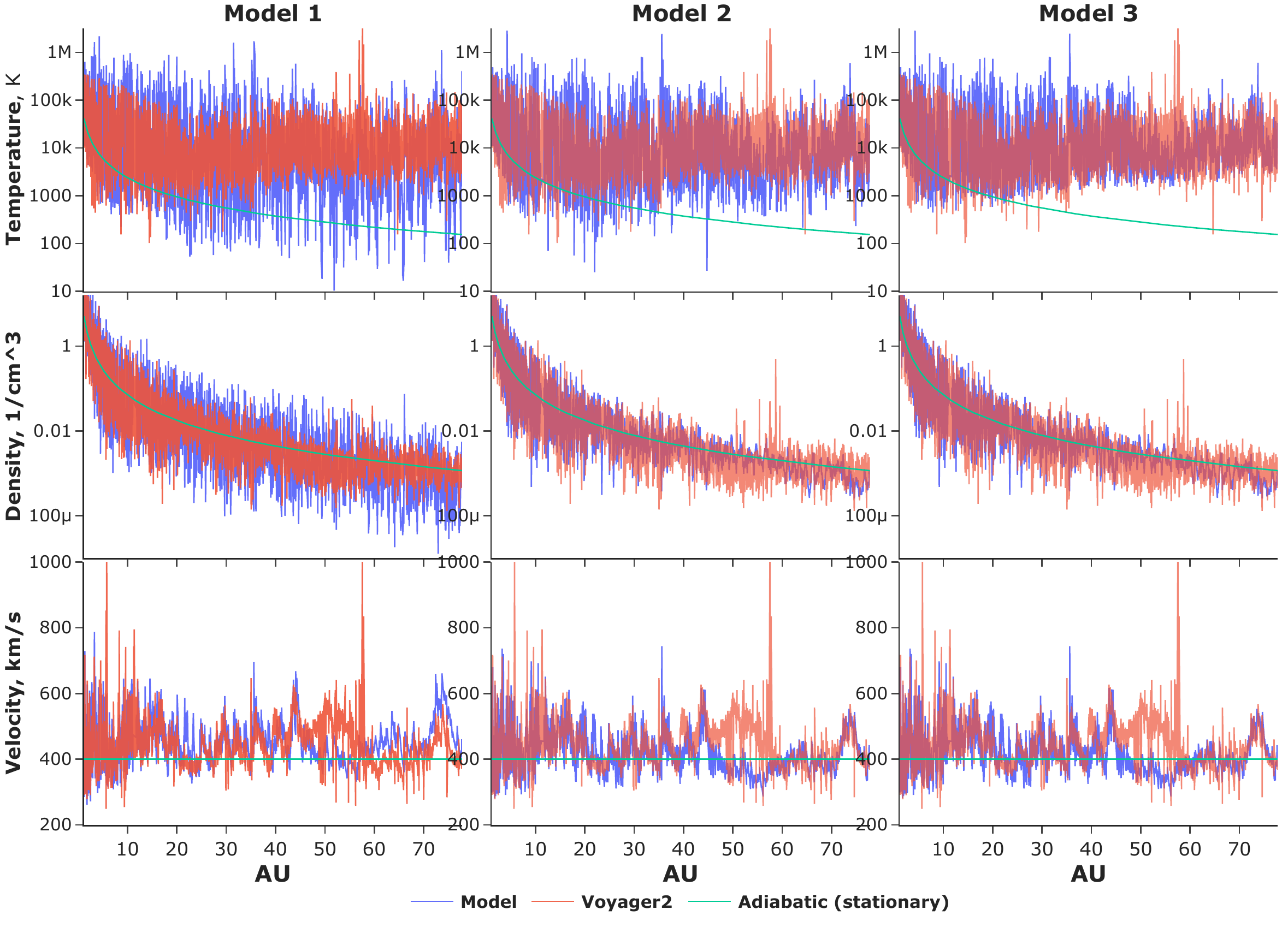}
	\caption{Temperature, density, and velocity distributions as functions of radial distance calculated along Voyager 2 trajectory. In contrast with Fig. \ref{fig1}, the results are not averaged. Blue curves correspond to modeling results, while red curves present Voyager data. The green curves are the stationary adiabatic solutions.  The value $n_H = 0.05\  \text{1/cm}^3$ has been used for Models 2 and 3.}
	\label{fig2}
\end{figure*}

In this Appendix, we present two figures that were not included in the main text but may be of interest to the reader.
Figure~\ref{fig2} presents the results of the modeling calculations without additional time averaging, as was done for Fig.~\ref{fig1}. It is interesting to note that Model 3 reproduces the temperature minima better than other models.

\begin{figure*}
	\includegraphics[width=0.99\hsize]{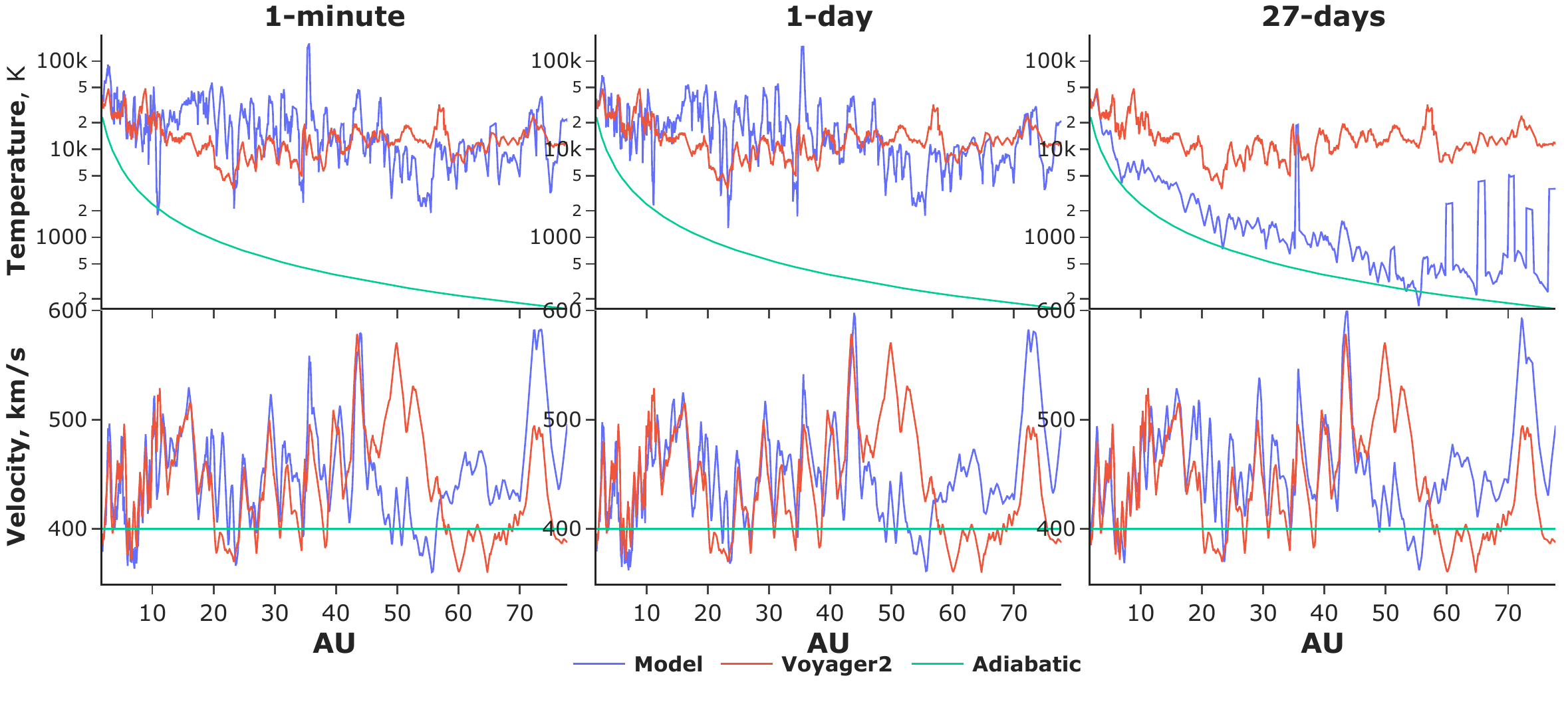}
	\caption{Temperature and velocity distributions obtained in the framework of Model 1 with different time averaging of OMNI data as the inner boundary conditions at 1 AU.  The left column presents calculations with one-minute OMNI data, the central column shows the same but with one-day-averaged OMNI data, and the right column the same but with 27d-averaged OMNI data.   Blue curves show results of the model calculations. Red curves present Voyager data. Green curves present the stationary adiabatic solution.}
	\label{fig3}
\end{figure*}

Figure~\ref{fig3} demonstrates the dependence of the numerical results on the time resolution of the inner boundary conditions.
The figure presents the results of numerical calculations in the frame of Model 1 with the 1min, 1d, and 27d averaged OMNI data at 1 AU.
It is clearly seen that the results of calculations with a 1min and 1d time resolution are almost identical, while the distribution with 27d-averaged boundary conditions differs significantly. The latter solution is somewhat close to the adiabatic solution.  It can be concluded that the 27d averaging in the inner boundary produces an overly smooth and unrealistic solution. The number of shock waves is much lower in such a smooth model and therefore the effect of shock-wave heating is not produced.
All models with the higher time resolution produce the shock heating effect rather well.

\label{ap3}
\label{pril1}

\section{Shock wave capture}
\label{ap4}
Numerous observations of the solar wind at 1 AU show that the shock-wave structure passes the observer within a time-frame of less than one minute. One might therefore wonder how numerical results with daily averaged boundary conditions can produce the effect of shock heating. To answer this question, we performed a numerical demonstration and found it very interesting and instructive.
 First of all, we chose (quite arbitrarily) a shock wave at 1 AU.  To do that, we used a database of the solar wind discontinuities prepared by \cite{Ermolaev2021} from OMNI dataset (\url{https://omniweb.gsfc.nasa.gov/}). We then extended this data to long distances using Model 1 and compared the results depending on the resolution of the boundary conditions (minute and daily).

\begin{figure*}
	\includegraphics[width=0.94\hsize]{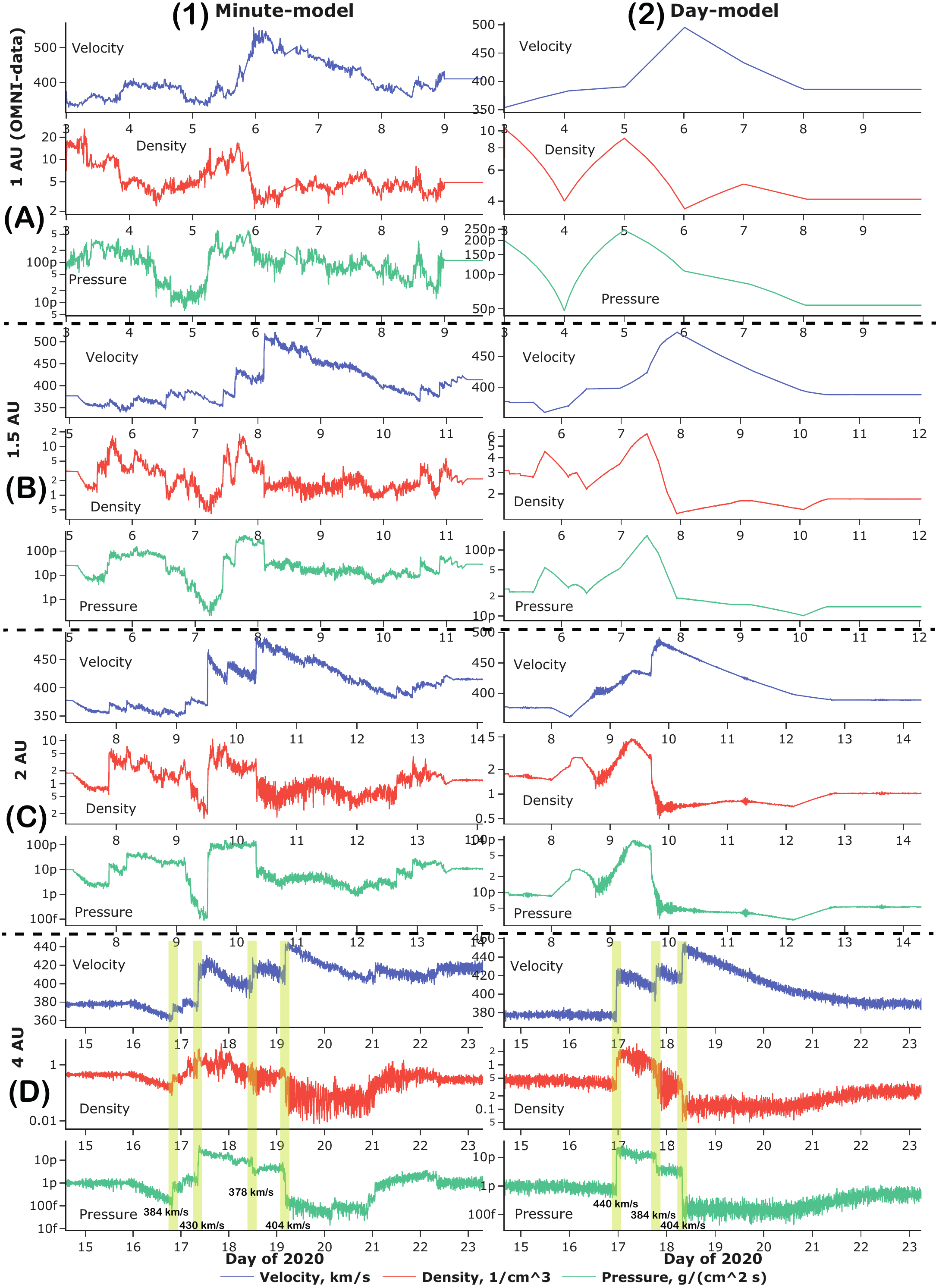}
	\caption{Solar wind speed, density, and pressure measurements at 1 AU from ONMIWeb (A) and the result of the  model calculations at 1.5 AU, 2 AU, and 4 AU (B-D, respectively). The two columns correspond to the models with one-minute (1) and one-day (2) resolutions of the boundary conditions. The x-axis is in days of 2020. The symbol <<p>>  denotes $10^{-12}$ and <<f>> is $10^{-15}$}
	\label{fig4}
\end{figure*}

 
Figure~\ref{fig4} shows two columns corresponding to the two-model runs with one-minute (1) and one-day (2) boundary conditions. The first series of calculations (A) shows measurements of the solar wind velocity, density, and pressure at 1 AU with one-minute (1A) and one-day (2A) resolutions. We selected 6 days of data from January 3, 2020, to January 8, 2020. This period includes the shock (Jan 5, 2020) as it is seen in the 1min data (1A). The next three series correspond to the speed, density, and pressure obtained in the models at 1.5 AU, 2 AU, and 4 AU respectively. The x-axis represents the days from the beginning of 2020.
 
The daily averaged data have only six points and do not catch the shock at 1 AU. However, velocity, pressure, and density variations still exist in the daily averaged data (as seen in panel 2A). Due to these variations, the faster solar wind overcomes the slower and the pressure wave is created and moves out. The pressure wave is overturned at some distance (between 1.5 and 2 AU as it is seen from panels 2B and 2C) and a shock or several shocks are created. It is seen undoubtedly (by comparing panels 1D and 2D) that the shock wave structures obtained in the runs with a minute and daily data are qualitatively similar. Although some time lag exists in the model with daily data. Using numerically obtained values of gas dynamical parameters on both sides of the shock and the shock speed, we verified that Rankine-Hugoniot conditions are satisfied. The shock-wave velocities are also shown in the figure (see 1D and 2D).

In this regard, the daily averaged boundary conditions at 1 AU  do not catch shocks at 1 AU. However, the shock structures are formed at a few AU even with a daily resolution of the boundary conditions. This explains why we observed the shock-heating mechanism in the runs with daily data.
 
\end{appendix}
\end{document}